\documentclass[cits]{PoS}
\usepackage{graphicx}
\usepackage{amsmath,amssymb,amsfonts,graphicx}
\usepackage[svgnames]{xcolor}

\newcommand{\vect}[1]{\textbf{\textit{#1}}}%ak

\newcommand{\Eq}[1]{Eq.~(\ref{#1})}
\newcommand{\al}[1]{\begin{align} #1 \end{align}}

\def\SA{\sphericalangle}

\def\vf{\varphi}
\def\fh{\varphi_h}

\def\fS{\varphi_S}
\def\fR{\varphi_R}
\def\fRB{\varphi_{\bar{R}}}

\def \fk{\varphi_{k}}

\def\fKR{\varphi_{KR}}

\def \fq{\varphi_{q}}
\def \fqR{\varphi_{qR}}
\def \fqRB{\varphi_{q\bar{R}}}
\newcommand{\lLa}{s_{L}}

\newcommand{\llangle}{\left\langle}
\newcommand{\rrangle}{\right\rangle}
\def\kT{\vect{k}_T}
\def\kBT{\bar{\vect{k}}_T}
\def\RT{\vect{R}_T}

\def\RBT{\bar{\vect{R}}_T}
\def\qT{\vect{q}_T}
\def\PT{\vect{P}_T}

\def\pT{\vect{p}_T}

\def\PT{\vect{P}_T}

\title{New way to access the quark fragmentation functions in electron-positron annihilation}

\ShortTitle{New way to access the quark fragmentation functions in electron-positron annihilation}

\author{\speaker{Aram~Kotzinian}
\thanks{ORCID: http://orcid.org/0000-0001-8326-3284}\\
AANSL (YerPhI), Yerevan, Armenia and Turin Division of INFN, Turin, Italy\\
E-mail: \email{aram.kotzinian@cern.ch}}
\author{Hrayr~H.~Matevosyan
\thanks{ORCID: http://orcid.org/0000-0002-4074-7411}\\
CSSM, Department of Physics, \\
The University of Adelaide, Adelaide SA 5005, Australia
\\ http://www.physics.adelaide.edu.au/cssm
}
\author{Anthony~W.~Thomas
\thanks{ORCID: http://orcid.org/0000-0003-0026-499X}\\
ARC Centre of Excellence for Particle Physics at the Terascale,\\     
and CSSM, Department of Physics, \\
The University of Adelaide, Adelaide SA 5005, Australia
}

\abstract{The description of the polarized quark hadronization process is one of the most challenging problems in the physics of strong interactions. The various single hadron and dihadron fragmentation functions, that quantify this process, are determined by analyzing the inclusive production of hadrons in electron-positron semi-inclusive annihilation process (SIA). These, in turn, are used to extract the transverse momentum dependent parton distribution functions from the experiments on semi-inclusive deep inelastic scattering process (SIDIS), elucidating  the spin-orbit correlations in nucleon. 
 
 Here we present a framework for new measurements of polarized quark fragmentation functions in electron-positron annihilation and deep inelastic semi-inclusive processes. These measurements offer a number of exciting opportunities to improve our understanding of the polarized quark hadronization and test the universality of the fragmentation functions. Such measurements can be performed at the upcoming {\tt BELLE~II}, {\tt JLab~12} and the future {\tt EIC} experiments.
}

\FullConference{XXVII International Workshop on Deep-Inelastic Scattering and Related Subjects - DIS2019\\
		8-12 April, 2019\\
		Torino, Italy}

\begin{document}

\section{Introduction}
The understanding of the internal structure of the nucleon remains one of the most intriguing issues of the modern physics. At high energies, the semi-inclusive reactions with large momentum transfer are described within the framework of perturbative quantum chromodynamics (pQCD) by convolutions of the transverse momentum dependent (TMD) parton distribution functions (PDFs) and fragmentation functions (FFs). Thus, to have an access to TMD PDFs and study their quark flavor and kinematic dependences, the knowledge of FFs and their flavor dependence is needed. Recently, we have examined the modeling and the phenomenology of one and two hadron TMD FFs in the series of articles~\cite{Matevosyan:2013eia,Matevosyan:2016fwi,Matevosyan:2017alv,Matevosyan:2017uls,Matevosyan:2017liq,Matevosyan:2018icf,Matevosyan:2018jht}.

 The measurements of two-hadron azimuthal asymmetries in semi-inclusive process have been recently used to access the quark transversity PDF of the nucleon~\cite{Bacchetta:2011ip,Bacchetta:2012ty,Radici:2015mwa}, using a combined analysis of two-hadron production in SIDIS and back-to-back two hadron pair creation in $e^+e^-$ annihilation. A key role here is played by  the so-called interference DiFF (IFF), that describes a correlation between the transverse spin polarization of a fragmenting quark with the relative transverse momentum of the produced hadron pair. Similarly, the helicity-dependent  DiFF, $G_1^\perp$,  describes a correlation between the longitudinal polarization of a fragmenting quark and the relative transverse momenta of the pair of hadrons. The importance of $G_1^\perp$ lies, to a considerable extent, in its relationship to the phenomenon of longitudinal jet handedness, which was predicted 25 years ago~\cite{Efremov:1992pe}, but has not yet been observed experimentally. It is also of interest because it has no analogue in single unpolarized hadron production. 

In general, the expressions for the cross-sections of SIDIS and SIA contain many terms with spin (in)dependent azimuthal modulations. In the leading order of pQCD formalism these terms contain complicated transverse momentum convolutions of TMD PDFs and FFs. The idea of using transverse-momentum weighting to break up the transverse momentum convolutions and to single out the PDFs and FFs in the single hadron azimuthal asymmetries was first proposed a number of years ago in~\cite{Kotzinian:1995cz,Kotzinian:1997wt} and later in~\cite{Boer:1997nt}. The first experimental results motivated by that work have recently been published by the {\tt COMPASS} collaboration~\cite{Alexeev:2018zvl,Matousek:2017xpc}. 

In this manuscript we present our recent results for the calculations of various weighted asymmetries of dihadron production in SIDIS and SIA. These weighted asymmetries allow us to disentangle the transverse momentum convolutions between the various PDFs and FFs entering the corresponding expressions, and allow us to study the various DiFFs.

\vspace{-0.2cm} 
\section{The weighted asymmetries for 2h+2h production in SIA}
\label{SEC_2h2h_SIA}
\vspace{-0.2cm}

  The first proposed azimuthal asymmetry for measuring  helicity-dependent DiFFs $G_1^\perp$, in back-to-back  two hadron pair creation in $e^++e^-\to (h_1 h_2) + (\bar{h}_1  \bar{h}_2) +X$, was made over ten years ago in~\cite{Boer:2003ya}. A subsequent experimental search for this asymmetry by BELLE collaboration did not yield a signal~\cite{Abdesselam:2015nxn, Vossen:2015znm}, while their previous measurements for IFF signal were sizable~\cite{Vossen:2011fk}. The model calculations of $G_1^\perp$ entering this asymmetry was recently calculated in~\cite{Matevosyan:2017alv}, producing a result that is smaller than that for IFF calculated within the same model, but still non-negligible.	
	  
  Recently, motivated by the findings in~\cite{Matevosyan:2017uls}, we re-derived the cross section for dihadron production in  $e^+e^-$ annihilation~\cite{Matevosyan:2018icf}, and found a number of disagreements with the previous calculations. The two most important conclusions were the resolution of the apparent inconsistencies between the definitions of IFF entering the SIDIS and SIA processes and the realization that the originally proposed azimuthal asymmetry for  determining $G_1^\perp$ in $e^+e^-$ annihilation should vanish.
	
\begin{figure}[t]
\centering 
\includegraphics[width=0.6\columnwidth]{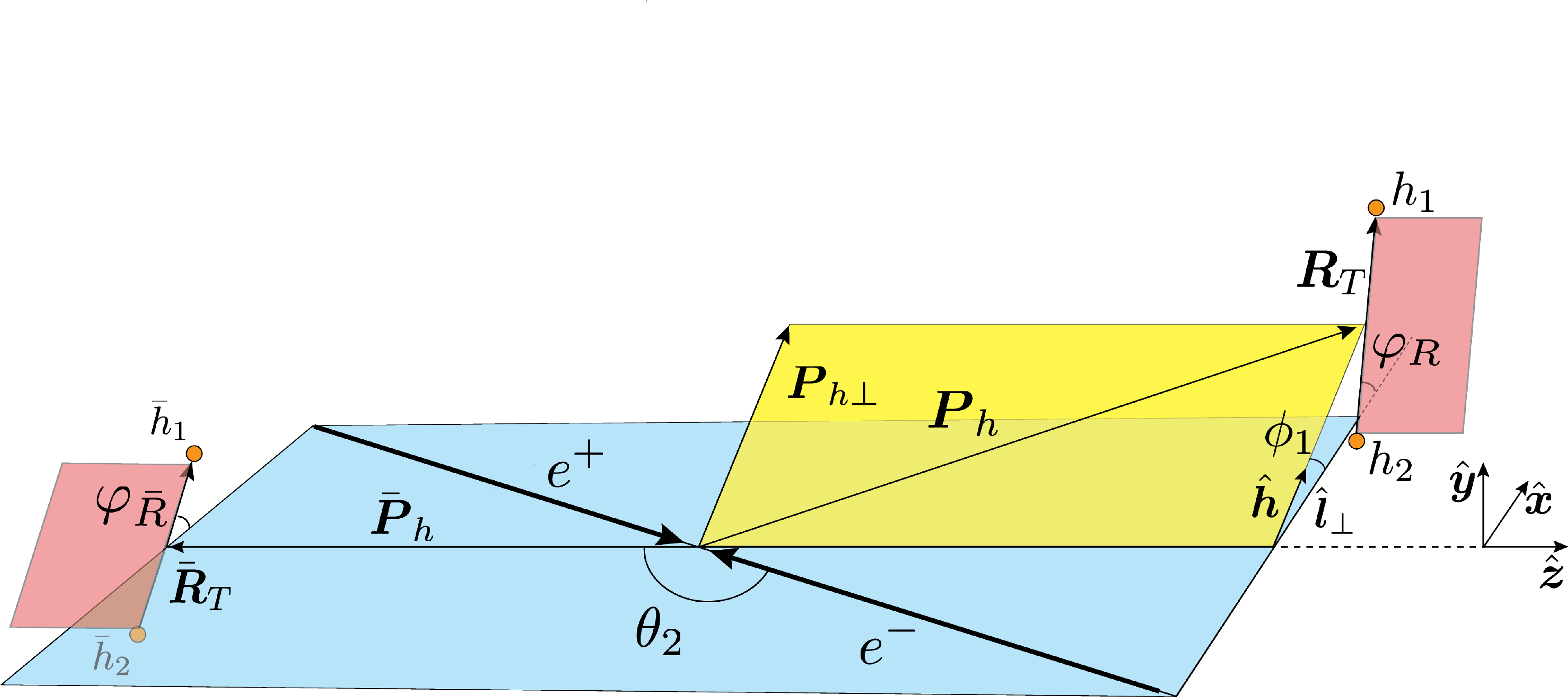}
%\GapCapt
\caption{The kinematics of two back-to-back dihadron pair creation in $e^+e^-$ annihilation.}%
\label{PLOT_EE_KINEMATICS-2h_2h}
\vspace{-0.1cm}
\end{figure}

 The coordinate system used in the analysis is defined in the center of mass frame of colliding $e^+$ and $e^-$ by taking the $\hat{z}$ opposite to the total 3-momentum ${\bar{\vect P}}_h$ of the hadron pair ($h_1h_2$), and the components of the vectors perpendicular to $\hat{z}$ are denoted with a subscript $_\perp$,  see Fig.~\ref{PLOT_EE_KINEMATICS-2h_2h}. In this frame, the perpendicular component of the virtual photon's  three-momentum ${\vect q}_\perp=0$. It is also useful to define a reference frame where the total momenta of both hadron pairs are collinear~\cite{Boer:2003ya}, and the components of 3-vectors perpendicular to them is denoted by $_T$. In this frame the virtual photon has a transverse momentum component ${\vect q}_T = -{{\vect P}}_{h\perp} /{z}$, where $z$ is the light-cone momentum fraction of $P_h$ with respect to $q$. The difference between the $_T$ and $_\perp$ components of the observed hadron momenta is  of the order $1/Q$ and will be neglected here, where $Q^2 = q^2$ is much larger than the typical hadronic scale. Note, that the azimuthal angle of $\qT$ and $\PT$ are related as $\fq= \vf_h+ \pi$.

%===============================================================================
%===============================================================================

 The cross section expression for this process at leading twist was originally derived in~\cite{Boer:2003ya}. The new derivation, in~\cite{Matevosyan:2018icf}, corrected several errors in the previous result. The  part of the cross section giving access to $G_1^{\perp a}$ is given by
\vspace{-0.3cm}
\al
{
\label{EQ_XSEC}
\hspace{-4.mm}\frac{d \sigma_{U,L}^{e^+e^- \to (h_1 h_2) (\bar{h}_1 \bar{h}_2) X }}{d^2 \qT \ d\fR\  d\fRB
\ d^7 V 
} 
 = C\sum_{a,\bar{a}} e_a^2  z^2 \bar{z}^2  A(y)
% 
%\\\non
%&\times
\Big\{
\mathcal{F}\Big[ D_1^a \bar{D}_1^{\bar{a}}\Big]
- \mathcal{F}\Big[\frac{(\RT\times \kT)_3}{M_h^2} \frac{(\RBT \times \kBT)_3}{\bar{M}_h^2}  G_1^{\perp a} \bar{G}_1^{\perp \bar{a}}  \Big]
 \Big\},
}
where $D_1^a$ denotes the unpolarized DiFF, $C\equiv\frac{3 \alpha^2}{\pi Q^2}$, while $\fR, \fRB$ are the azimuthal angles of $\RT$ and $\RBT$, and $d^7V\equiv dz d\xi d R_T d \bar{z} d\bar{\xi}  d \bar{R}_T dy$ denotes the remaining phase space element. The subscript $3$ denotes the $z$ component of the vector, and $M_h$ and $\bar{M}_h$ are the invariant masses of the two hadron pairs. The electromagnetic coupling constant is denoted  as $\alpha$, while the charge of a quark of flavor $a$ is $e_a$, and 
{
$A(y) = \frac{1}{2} - y + y^2$.
}
The convolution $\mathcal{F}$ is defined as
\vspace{-0.3cm}
\al
{
\label{EQ_F_CONVOL}
\mathcal{F}[w D^a \bar{D}^{\bar{a}} ]
 =  \int d^2 \kT &d^2 \kBT 
 \  \delta^2(\vect{k}_T + \bar{\vect{k}}_T - \vect{q}_T)
%\\ \non
% & \times 
  w( \kT, \kBT, \RT, \RBT)\ D^a \ D^{\bar{a}}.
}

The fully unintegrated DiFFs entering these expressions depend on the relative azimuthal angles between $\kT$, $\RT$ and  $\kBT$, $\RBT$, respectively: $D^a(z, \xi, \kT^2, \RT^2, \kT \cdot \RT  )$, $D^{\bar{a}}(\bar{z}, \bar{\xi}, \kBT^2, \RBT^2, \kBT \cdot \RBT  )$.

In order to gain information about the DiFFs entering the cross section, it is helpful to use their decomposition into Fourier cosine series with respect to the relative azimuthal angle $\fKR \equiv \fk - \fR$
\al
{
\label{EQ_D_FOURIER}
D^a(z, \xi, \kT^2, \RT^2,& \cos(\fKR)) 
%%
%\\ \non
%  =& 
	=\frac{1}{\pi} \sum_{n=0}^\infty
 \frac{\cos(n \cdot \fKR)}{1+\delta^{0,n}} \ D^{a,[n]}(z, \xi, \kT^2, \RT^2),
}

The cross section, when integrated over $\qT, \fR, \fRB, \xi, \bar{\xi}$, contains only the unpolarized DiFFs
\vspace{-0.3cm}
\al
{
\label{EQ_XSEC_INT_UPOL}
 \langle 1 \rangle %(y, z, M_h^2, \bar{z}, \bar{M}_h^2)
 &=  \int d \sigma_{U,L}^{e^+e^- \to (h_1 h_2) (\bar{h}_1 \bar{h}_2) X }\  \times 1
 %
% \\ \non
%& 
=   C  A(y)
 \sum_{a,\bar{a}} e_a^2   D_1^a(z, M_h^2) \bar{D}_1^{\bar{a}}(\bar{z}, \bar{M}_h^2),
}
where the integrated zeroth Fourier moment of the unpolarized DiFF is defined as
\vspace{-0.3cm}
\al
{
D_1^{a}(z, M_h^2)  =  z^2 \int d^2 \kT \int  d \xi
%  
%\\ \non
%&\times  
\ D_1^{a,[0]}(z, \xi, \kT^2, \RT^2) \, .
}
%
%Here $\RT^2$ has been substituted by the invariant mass square, $M_h^2$, of the the hadron pair, per convention. 

It is easy to derive from \Eq{EQ_XSEC}, that the previously proposed measurement of an asymmetry containing $G_1^\perp$ in~\cite{Boer:2003ya} simply vanishes
\vspace{-0.3cm}
\al
{
\langle \cos(2(\fR - \fRB)) \rangle =0.
}
Moreover, it is easy to demonstrate, that $\langle f(\fR, \fRB)\rangle=0$ for an arbitrary $f(\fR, \fRB)$.

In~\cite{Matevosyan:2018icf} we proposed a new weight to access the helicity-dependent DiFFs
\vspace{-0.2cm}
\al
{
\label{EQ_PHI1_AV}
&\left\langle \frac{q_T^2
\big(3 \sin(\fqR) \sin(\fqRB) 
+ \cos(\fqR) \cos(\fqRB) \big)
}{M_h \bar{M}_h  }
 \right \rangle
= 4C 
%\ 
 \sum_{a,\bar{a}}  e_a^2  
 \Big( G_1^{\perp a,[0]}  - G_1^{\perp a,[2]} \Big) \Big( \bar{G}_1^{\perp \bar{a}, [0]}  - {G}_1^{\perp \bar{a}, [2]} \Big),
}
where the dimensionless integrated $n$-th moments are defined as
\vspace{-0.3cm}
\al
{
\label{EQ_G_MOMS}
G_1^{\perp a,[n]}(z, M_h^2) \equiv  &\ z^2  \int d^2 \kT  \int  d \xi 
%\\ \non
%&\times
 \left(\frac{\kT^2}{2 M_h^2}\right)  \frac{|\RT|}{M_h} \ G_1^{\perp a,[n]}(z, \xi, \kT^2, \RT^2).
}
For convenience, we  define the integrated helicity-dependent DiFF as
\al
{
\label{EQ_G1_INT}
G_1^{\perp a}(z, M_h^2) \equiv  G_1^{\perp a,[0]}(z, M_h^2)  - G_1^{\perp a,[2]}(z, M_h^2).
}
Note, that this definition differs from that in~\cite{Boer:2003ya}. Our recent model calculations of Fourier cosine moments of $G_1^\perp$ in~\cite{Matevosyan:2017alv} suggest a sizable analyzing power for such combination, although in the model calculations we used $k_T R_T$ wighting when defining the Fourier cosine moments of $G_1^\perp$ instead of $k_T^2 R_T$ used in \Eq{EQ_G_MOMS}.

The corresponding azimuthal asymmetry, which is the ratio of the weighted moment in \Eq{EQ_PHI1_AV} to the unweighted one in \Eq{EQ_XSEC_INT_UPOL}, can be expressed as
\vspace{-0.3cm}
\al
{
\label{EQ_ee_SSA}
 A_{e^+e^-}^{\Rightarrow}(z ,& \bar{z}, M_h^2, \bar{M}_h^2)
% \\ \non
 = 4
\frac{ \sum_{a,\bar{a}} 
 G_1^{\perp a}(z, M_h^2) \ G_1^{\perp \bar{a}}(\bar{z}, \bar{M}_{h}^2)
 }
 { \sum_{a,\bar{a}} 
 D_1^a(z, M_h^2) \ D_1^{\bar{a}} (\bar{z}, \bar{M}_{h}^2) 
 }.
}
%

%%%%%%%%%%%%%%%%%%%%%%%%%%%%%%%%%%%%%%%%%%%%%%%%%%%%%
%%%%%%%%%%%%%%%%%%%%%%%%%%%%%%%%%%%%%%%%%%%%%%%%%%%%%
%%%%%%%%%%%%%%%%%%%%%  SECTION  %%%%%%%%%%%%%%%%%%%%%%%%%%
\section{The weighted asymmetries for 2h production in SIDIS}
\vspace{-0.2cm}

We next consider the SIDIS process with two observed final state hadrons $l + N \to l' + h_1h_2 +X$. For this process, using the the expression for the fully unintegrated cross section of this process, derived in~\cite{Bacchetta:2002ux}, we suggested a new weighted asymmetry for accessing $G_1^\perp$ in~\cite{Matevosyan:2018icf} 

The cross section for this process can be decomposed into various terms according to the polarization of the incident lepton beam and the target nucleon. The two cases of interest here are $\sigma_{UU}$ and $\sigma_{UL}$, describing the unpolarized and target longitudinal polarization dependent parts of the cross section, respectively. Here we only show the  explicit dependence of these two terms on the relevant azimuthal angles  
\vspace{-0.2cm}
\al
{
\label{EQ_SIDIS_XSEC_UU}
\frac{d \sigma_{UU} }{ d^2 \vect{P}_{h\perp} d\fR  \ d^6V'}
=  C'A'(y)\sum_a e_a^2  \
 \mathcal{G} \Bigg[ f_{1}^a\ D_1^{ a}
 \Bigg],
}
and
\vspace{-0.5cm}
\al
{
\label{EQ_SIDIS_XSEC_OL}
&\frac{d \sigma_{UL} }{ d^2 \vect{P}_{h\perp} d\fR \ d^6V'}
%
%\\\non
%&\hspace{1cm}
= -  S_L \sum_a \frac{\alpha^2 e_a^2}{\pi y Q^2}  A'(y)
 \mathcal{G} \Bigg[\frac{(\RT \times \kT)_3}{M_h^2} g_{1L}^a  G_1^{\perp a} \Bigg],
}
where $C'=\frac{\alpha^2 }{\pi y Q^2}$, $d^6V' \equiv dz d\xi dM_h^2 dx dy d\fS$, $\fS$ is the azimuthal angle of the initial  nucleon's transverse polarization $\vect{S}_T$, $A'(y) = 1- y + y^2/2$, $S_L$ is the longitudinal polarization of the nucleon, and the SIDIS convolution is defined as
\al
{
\label{EQ_G_CONVOL}
\mathcal{G}[&w f^q D^{q} ] \equiv \int d^2 \pT  \int d^2 \kT
   \delta^2\Big(\kT -\pT+ \frac{\vect{P}_{h\perp}}{z} \Big)
%\\ \non   
%&\times    
w( \pT, \kT, \RT)
f^q(x, p_T) D^{q}(z, \xi, k_T^2, R_T^2, \kT \cdot \RT  ).
}

The unweighted integral of the cross section over $ \vect{P}_{h\perp}, \fR,\xi, \fS$ yields a product of the unpolarized PDF and the DiFF
\vspace{-0.1cm}
\al
{
& \left\langle 1 \right\rangle 
 = \int d \sigma_{UU} \times 1
=  2C'A'(y)\sum_a e_a^2 f_1^a(x) z^2 D_1^a (z, M_h^2).
}

For the polarization-dependent term in \Eq{EQ_SIDIS_XSEC_OL}, again by weighting this modulation by a factor of $P_{h\perp}$, we can break up the transverse momentum  convolution in~(\ref{EQ_G_CONVOL}) into a product of two independent terms
\al
{
 \left\langle \frac{P_{h\perp} \sin(\fh - \fR)}{M_h} \right\rangle 
 &= \int d \sigma_{UL}    \frac{P_{h\perp} \sin(\fh - \fR)}{M_h}
%
%\\ \non
%&\hspace{-3.2cm}
= 2 S_L C' A'(y)\sum_a  
\  g_{1L}^a(x) \ z\ G_1^{\perp a}(z, M_h^2),
}
where
\vspace{-0.5cm}
\al
{
 g_{1L}^a(x) = \int d^2 \pT \ g_{1L}^a(x, p_T^2),
}
is the nucleon collinear helicity PDF, while the combination of the Fourier cosine moments of  $G_1^{\perp a}(z, M_h^2)$ is exactly that appearing in the $e^+e^-$ annihilation asymmetry in \Eq{EQ_G1_INT}.

Thus, the proposed azimuthal asymmetry can be expressed as 
\al
{
\label{EQ_SIDIS_SSA}
 A^{\Rightarrow}_{SIDIS}(x, z , M_h^2)
 =&S_L \frac{\sum_{a}  g_{1L}^a(x)
 \ z\ G_1^{\perp a}(z, M_h^2) }
 { \sum_{a} 
f_1^a(x)\ D_1^a(z, M_h^2)  }.
}
%

%===============================================================================
%%%%%%%%%%%%%%%%%%%%%%%%%%%%%%%%%%%%%%%%%%%%%%%%%%%%%
%%%%%%%%%%%%%%%%%%%%%   SUB SECTION %%%%%%%%%%%%%%%%%%%%%%%
\section{The weighted asymmetries for h+2h production in SIA}
\vspace{-0.2cm}

Recently we proposed a new measurement in $e^+e^-$ annihilation~\cite{Matevosyan:2018jht}, exploiting the production of a single inclusive hadron (generically noted as $\Lambda$) back-to-back to a hadron pair, where the relevant cross section should involve convolutions of FFs for the single hadron in one jet and the DiFFs for the hadron pair produced in the opposite jet. The purpose of such a measurement is two-fold, and will leverage our knowledge of the single hadron FFs. First, the absolute cross section measurements will provide a wider basis for extracting the quark flavor dependence of the DiFFs, especially when analyzed together with the inclusive hadron pair measurements in the same jet~\cite{Seidl:2017qhp}. Second, by studying various azimuthal asymmetries we can better determine the polarized DiFFs, and also access their sign.

%===============================================================================
\begin{figure}[htb]
\centering 
\includegraphics[width=0.6\columnwidth]{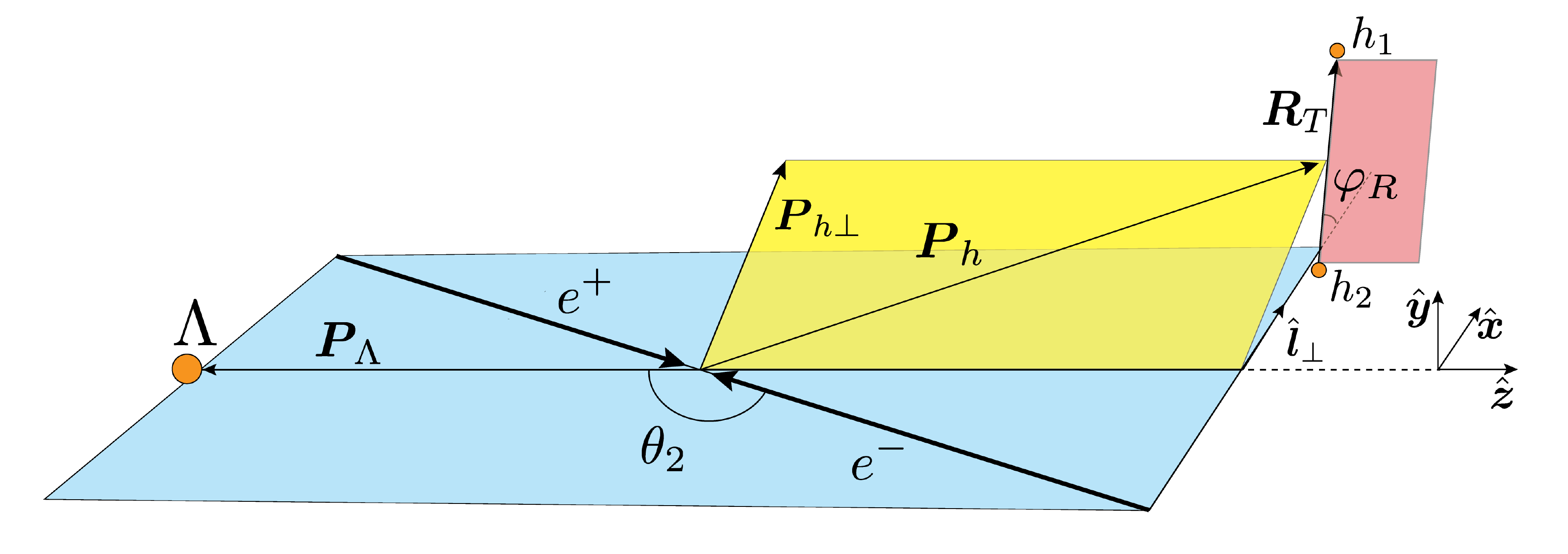}
\caption{The kinematics of $e^+e^-\to h_1 h_2 +\Lambda + X$ process.}
\label{PLOT_EE_KINEMATICS-2h_h}
\end{figure}
%===============================================================================

  Here we use the $e^+e^-$ center-of-mass coordinate system, similar to that  in Sec.~\ref{SEC_2h2h_SIA}, where the $\hat{z}$ axis is chosen to point opposite to the $\Lambda$'s 3-momentum $\vect{P}_\Lambda$. Some of the results for weighted asymmetries from~\cite{Matevosyan:2018jht} are:

\hspace{-0.8cm}
1. The weighted asymmetry induced by the Collins effect for unpolarized hadron $\Lambda$:
\vspace{-0.3cm}
\al
{
\label{EQ_ASYMM_COLL}
 A^{Coll} =  \frac{B(y)}{A(y)} \dfrac{\sum_{a} e_a^2\ H_1^{\SA, a\to h_1 h_2}(z, M_h^2) \  H_{1}^{\perp \bar{a}, [1]}(\bar{z}) } {\sum_{a} e_a^2\ D_1^{a\to h_1 h_2}(z, M_h^2) \ \bar{D}_1^{\bar{a} \to \Lambda}(\bar{z}) } \, ,
}
where $B(y)=y(1-y)$.

\hspace{-0.8cm}
2. The acquired longitudinal polarization $\lLa$ of $\Lambda$:

\vspace{-0.7cm}
\al
{
\label{EQ_POL_G1}
 \llangle \lLa\rrangle^{\sin(\fq - \fR)} (z, M_h^2,\bar{z},y) = \dfrac{\sum_{a} e_a^2 
\ G_1^{\perp,a\to h_1 h_2}(z, M_h^2) \  G_{1L}^{\bar{a}\to \Lambda}(\bar{z})
}
{
\sum_{a} e_a^2  
\ D_1^{a\to h_1 h_2}(z, M_h^2) \ \bar{D}_1^{\bar{a} \to \Lambda}(\bar{z})
 },
 }
3. An asymmetry induced by the correlation of the transverse polarizations of the quark and the anti quark, which couples the Collin DiFFs with the $H_{1L}^{\perp }$ type FF:
\vspace{-0.3cm}
\al
{
\label{EQ_AV_H1L}
\llangle \beta_L \rrangle _{H_1^{\SA} H_{1L}^{\perp}} =  \llangle \frac{q_T}{M_\Lambda}\sin(\fq + \fR) \rrangle = \frac{3 \alpha_{em}^2}{(2 \pi)^2 Q^2} B(y) \sum_{a} e_a^2 
\ H_1^{\SA, a\to h_1 h_2}(z, M_h^2) \  H_{1L}^{\perp \bar{a}, [1]}(\bar{z}),
}
4. The asymmetries for a transversely polarized $\Lambda$:
\al
{
\frac{\llangle s_{y} \rrangle^{\cos(\fq)}(z, M_h^2,\bar{z},y)
 - \llangle s_{x} \rrangle^{\sin(\fq)}(z, M_h^2,\bar{z},y) }
 {M_\Lambda}
 = 2 \dfrac{\sum_{a} e_a^2 
 D_1^{ a\to h_1 h_2}(z, M_h^2)  \ D_{1T}^{\perp \bar{a}, [1]}(\bar{z})
}
{
\sum_{a} e_a^2 
 D_1^{a\to h_1 h_2}(z, M_h^2) \ \bar{D}_1^{\bar{a} \to \Lambda}(\bar{z})
 },
}
\vspace{-0.4cm}
\al
{
\frac{ \llangle s_{y} \rrangle^{\cos(\fq)}(z, M_h^2,\bar{z},y)
 + \llangle s_{x} \rrangle^{\sin(\fq)}(z, M_h^2,\bar{z},y) }
 {M_h}
%&
%%
%\\ \non
 %= & 
=\frac{2B(y)}{A(y)} \dfrac{\sum_{a} e_a^2 
  H_1^{\perp, a\to h_1 h_2}(z, M_h^2) \  H_{1}^{\bar{a}\to\Lambda}(\bar{z})
}
{
\sum_{a} e_a^2 
 D_1^{a\to h_1 h_2}(z, M_h^2) \ \bar{D}_1^{\bar{a} \to \Lambda}(\bar{z})
 }.
}
\vspace{-0.4cm}
\al
{
\label{EQ_POL_HaH1}
 \llangle \vect{ s}_{T} \rrangle^{\sin(\fR)}_{x}(z, M_h^2,\bar{z},y) 
 =  \llangle \vect{ s}_{T} \rrangle^{\cos(\fR)}_{y}(z, M_h^2,\bar{z},y)
%
 %\\ \non
 %& = 
 =\frac{1}{2}\frac{B(y)}{A(y)}\dfrac{\sum_{a} e_a^2 
\ H_1^{\SA, a\to h_1 h_2}(z, M_h^2) \  H_{1}^{\bar{a}}(\bar{z})
}
{
\sum_{a} e_a^2 \
 D_1^{a\to h_1 h_2}(z, M_h^2) \ \bar{D}_1^{\bar{a} \to \Lambda}(\bar{z})
 }.
 }

\section{Conclusions}
\vspace{-0.2cm}

In our recent work, we proposed a new inclusive measurement, where an unpolarized hadron pair is detected back-to-back with an another hadron pair or a single hadron, that may or may not be polarized. A number of new exciting measurements were then discussed. For example: 

\textbf{a}) New measurements which will permit us to access the helicity-dependent DiFF, both in back-to-back two hadron pair production in $e^+e^-$ annihilation and in forward two hadron production in SIDIS. 

\textbf{b}) A new measurement of the unpolarized cross section will enable us to determine the flavor dependence of the DiFF using our knowledge of the ordinary unpolarized FFs, which is not possible with single hadron $e^+e^-$ measurements. The measurement of the hadron polarization dependent asymmetries gives access to a wide variety of combinations of polarized spin $1/2$ baryon FFs and polarized DiFFs.

%%%%%%%%%%%%%%%%%%%%%%%%%%%%%%%%%%%%%%%%%%%%%%%%%%%%%%%%%%%%%%%%%%%%%%%%%%%%%%%
\newpage
\bibliographystyle{JHEP}
\bibliography{fragment}

\end{document}